\documentclass[notitlepage]{revtex4-2}%\linenumbers
\usepackage{hyperref}

\usepackage[utf8]{inputenc}

\usepackage{mathtools}
\usepackage{mathrsfs}
\usepackage{bbm}
\usepackage{xcolor}
\usepackage{graphicx}
\usepackage{bm}
\usepackage{tikz-cd}
\usepackage{array, multirow}
\usetikzlibrary{calc,patterns, angles, quotes}
\usepackage{siunitx}
\usepackage{tkz-euclide}
\usepackage{subcaption}
\usepackage{pgfplots}
\usepackage{amsmath, amssymb}
\usepackage{upgreek}
\usepackage{tabularx,booktabs}
\usepackage{siunitx}
\usepackage{float}

\newcolumntype{Y}{>{\centering\arraybackslash}X}

\DeclareMathAlphabet{\mathpzc}{OT1}{pzc}{m}{it}

\usepackage[cal=boondox]{mathalfa}

\begin{document}

\title{Significance of the negative binomial distribution in multiplicity phenomena}

\date{April 11, 2023}

\author{S. V. Tezlaf}
\affiliation{Vienna, Austria | s.v.tezlaf@gmail.com}

\begin{abstract}
The negative binomial distribution (NBD) has been theorized to express a scale-invariant property of many-body systems and has been consistently shown to outperform other statistical models in both describing the multiplicity of quantum-scale events in particle collision experiments and predicting the prevalence of cosmological observables, such as the number of galaxies in a region of space. Despite its widespread applicability and empirical success in these contexts, a theoretical justification for the NBD from first principles has remained elusive for fifty years. The accuracy of the NBD in modeling hadronic, leptonic, and semileptonic processes is suggestive of a highly general principle, which is yet to be understood. This study demonstrates that a statistical event of the NBD can in fact be derived in a general context via the dynamical equations of a canonical ensemble of particles in Minkowski space. These results describe a fundamental feature of many-body systems that is consistent with data from the ALICE and ATLAS experiments and provides an explanation for the emergence of the NBD in these multiplicity observations. Two methods are used to derive this correspondence: the Feynman path integral and a hypersurface parametrization of a propagating ensemble.
\end{abstract}

\maketitle

%\section{}
The negative binomial distribution (NBD) has served as an especially accurate statistical model for a broad range of multiplicity observations in particle collision experiments, e.g., $p \bar p$, $hh$, $hA$, $AA$, $e^+e^-$ \cite{Grosse-Oetringhaus_2010,PhysRevD.99.094045,tarnowsky2013first,derrick1986study,Zborovsky2018} (See \cite{kittel2005soft} for an overview), and is argued to be a scale-invariant property of matter \cite{schaeffer1984determination, schaeffer1985probability}, providing the best fit for astronomical observations, where it predicts the number of galaxies in a region of space \cite{perez2021void,hurtado2017best,10.1093/mnras/254.2.247,hameeda2021generalized}. Although several theories have been proposed to explain the relevance of the NBD, the precise physical principles behind its occurrence in these observations are unknown \cite{Grosse-Oetringhaus_2010,PhysRevD.99.094045,PRASZALOWICZ2011566,fry2013void,10.1093/mnras/254.2.247}. This work reveals a fundamental significance to the NBD beyond previous considerations, by showing that the model arises naturally from first principles and in a highly general setting via the relativistic dynamics of a freely-propagating canonical ensemble. These results lead to an emergent, statistical description of relativistic ensembles that is implicit to the vacuum and intrinsically quantum-mechanical, providing an explanation for the distribution's relevance at modeling multiplicity phenomena, regardless of scale, consistent with measurements from the ATLAS and ALICE experiments at the LHC in both double and triple NBD fits of charged particle multiplicity data over a range of collision energies and pseudorapidities.

Early applications of the NBD in hadronic multiplicity distribution models were reported by Giovannini (1973)\cite{giovannini1973thermal}, Carruthers, Shih and Duong-van (1983)\cite{PhysRevD.28.663}, and the UA5 collaboration at the CERN $p \bar p$ collider (1985)\cite{alner1985multiplicity}, which accounted for the observed violation of Koba, Nielsen and Olesen (KNO) scaling that signaled an energy dependence in multiplicity observations. Various implementations of the NBD to describe particle collision experiments continues to be an active area of research. Following the initial application of the NBD in this context, Carruthers and Duong-van \cite{CARRUTHERS1983116}, soon applied their NBD model to describe the number of galaxies in Zwicky clusters.  A hierarchical theory was later developed by Schaeffer \cite{schaeffer1984determination, schaeffer1985probability} that extended correlations of vanishingly small masses to objects of arbitrary mass and size, thus justifying the NBD's effectiveness in modeling celestial observations. This hierarchical scaling has been verified many times, in a variety of astronomical samples (e.g.,  \cite{maurogordato1987void,vogeley1994voids,fry1989void,maurogordato1992large,bouchet1993moments,croton2007statistical,conroy2005deep2,tinker2008void}). The distribution function is used to study the probability of observing $n$ galaxies in a number of disconnected, volumetric cells, where the void probability, or zero-point correlation function, which relates all higher-order correlation functions, has received increasing attention as a tool in the field, for which the NBD has proven the most effective \cite{perez2021void,hurtado2017best,10.1093/mnras/254.2.247,hameeda2021generalized}. The model can be described as the probability of recording a given number of ``successes" (observing a galaxy) after a certain number of ``failures" (voids), where the probability of a failure depends on the density of the sample and how clustered it is \cite{croton20042df}.  Despite the model's effectiveness, as the scale-invariant theory behind these astronomical observations relies on the behavior of a microscopic phenomenon---for which no physical explanation has yet been proven---it remains equally wanting of a logical foundation \cite{fry2013void, 10.1093/mnras/254.2.247}.

\paragraph*{The Negative Binomial Distribution---} 
The NBD is a probability mass function, typically parametrized as
\begin{align}\label{eq: NBD}
	\Pr(n) = \frac{\Gamma{(n+k)}}{n!\Gamma{(k)}}(1-\langle p \rangle)^k\langle p \rangle^n \quad  \quad \text{or}  \quad \quad \Pr(n) = \frac{\Gamma{(n+k)}}{n!\Gamma{(k)}}\left(\frac{k}{k+\langle n \rangle}\right)^k\left(\frac{\langle n \rangle}{k+\langle n \rangle}\right)^n,
\end{align}
where $n$ is the random variable (i.e., number of ``successes"), and $k\in \mathbb{R}$ is the dispersion parameter (i.e., number of ``failures"). Similar to the binomial distribution (BD)---which calculates the probability that, given ${\mathcal{N}}$ trials, $n$ will succeed and $k$ will fail---the NBD determines the likelihood that, given $k$ trials \textit{necessarily} fail, some $n$ number of trials will ultimately succeed. The total number of trials ${\mathcal{N}}$ is known \textit{a priori} for the BD but \textit{a posteriori} for the NBD. The inverse statement applies to the parameter $k$. Averaged over many experiments, the mean number of trials $\langle \mathcal{N} \rangle$ and mean count $\langle n \rangle$ are related by the mean \textit{chance} $\langle p \rangle$, denoting the probability of a single success event---where $\langle p \rangle = \langle n \rangle / \langle \mathcal{N} \rangle$. The mean parameters are related to $k$ by the formula $\langle \mathcal{N} \rangle - \langle n \rangle = k$. As $k$ is a constant in the NBD, it is not denoted as an average. More details on the NBD are provided in Appx. \ref{sec: NBD}.

\begin{figure}[ht]
\centering
\begin{subfigure}[b]{0.488\textwidth}\label{fig: MD 8 TeV}
\centering
\includegraphics[width=\linewidth]{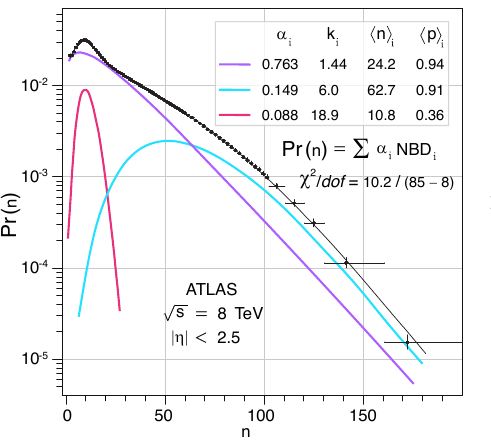}
\caption{}
\end{subfigure}
~\hfill%%%%%%%
\begin{subfigure}[b]{0.49\textwidth}
\centering
\includegraphics[width=\linewidth]{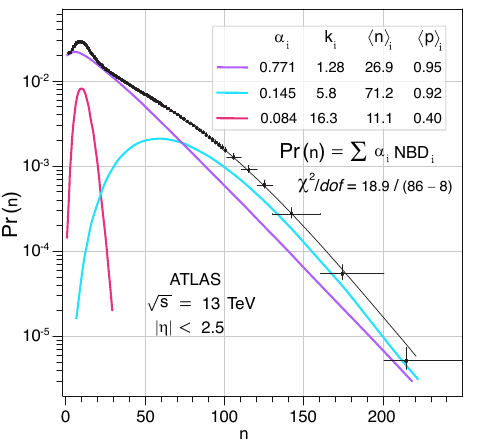}
\caption{} \label{fig: MD 13 TeV}
\end{subfigure}
           \caption{Multiplicity distributions of charged particles modeled by the superposition of three NBDs \cite{Zborovsky2018}, depicting data measured by the ATLAS Collaboration \cite{aad2016charged, aaboud2016charged} for events of $n>1$ in the pseudorapidity interval $|\eta| < 2.5$ with transverse momentum $\text{p}_T > 100$ MeV/c, (a) $\sqrt{s} = 8$ TeV and (b) $13$ TeV. Error bars include both the statistical and systematic uncertainties. The $\chi^2$ statistical methods are provided in the reference.} \label{fig: NBD MD} 
           \end{figure}

Although the origin of the NBD in multiplicity measurements of high-energy collisions is still unknown, some general qualitative features can be easily stated based on current observations. As the center-of-mass energy $\sqrt{s}$ increases, the average multiplicity of measured particles, modeled by $\langle n \rangle$, also increases, while the parameter $k$ decreases. In fact, $k^{-1}$ can be well-approximated by a linear function of $\ln{\sqrt{s}}$ \cite{Grosse-Oetringhaus_2010}. Moreover, the average multiplicity has been shown to increase with the pseudorapidity interval $|\eta|$, while the relationship between $k$ and $|\eta|$ exhibits a more complex behavior, varying among the different NBD models required to properly fit experimental data \cite{acharya2017charged}. Experiments at increasing energies have revealed structures in the multiplicity data that can no longer be described by a single NBD but rather require the superposition of multiple NBDs to accurately model observations. Initially, the requirement of two NBDs prompted the hypothesis that these distinct distributions might account for soft and semi-hard collision events \cite{giovannini1999possible, giovannini2005clan}. In more recent years, a third NBD has been found necessary to describe additional features of the data (\cite{zborovsky2013three, Zborovsky2018}, Fig. \ref{fig: NBD MD}), which is conjectured to describe yet another class of collision events. The precise physical significance of these additional NBDs is still being researched, and rather modest progress has been made toward these matters at the time of writing \cite{ Zborovsky2018}.

Interpretation of the NBD in terms of an underlying production mechanism that is apparently common to hadronic, leptonic, and semileptonic processes is still considered a challenging problem. Several theories have been developed to derive the NBD from general principles, including a stochastic model \cite{carruthers1983correlations}, a clan model \cite{giovannini1986negative}, a string model \cite{werner1989origin}, a stationary branching process \cite{chliapnikov1989negative}, a two-step model of binomial cluster production and decay \cite{iso1990negative}, and the glasma color flux tube model \cite{gelis2009glittering}. However, it is still not understood why the same distribution fits such diverse reactions and why the parameter $k$ decreases with increasing energy.

In these proceedings, we show that the NBD multiplicity model can indeed be derived from first principles and in an extremely general context, providing an explanation for the model's broad applicability without resorting to ad hoc mechanisms or speculative processes. These theoretical results are consistent with experimental observations and provide a promising foundation for generating more precise predictions through future refinements.

We first demonstrate the correspondence of a statistical event of the NBD to a spacetime event in Minkowski space by considering the observation of a canonical ensemble of massive, relativistic particles along the radial axis. After generalizing these results to three spatial dimensions, we will recover exactly the same conclusions from a completely different approach, via the quantum mechanical path integral.

\subsection*{Correspondence of Statistical/Spacetime Events}

\paragraph*{The NBD and hyperbolic parametrizations along the radial axis---}\label{sec: Correspondence of NBD and Hyperbolic 2-Space Parametrizations}
In a Minkowski spacetime with metric signature $(+, -, -, -)$, let a timelike event in an inertial frame $O$ be denoted ${x^\mu \equiv (x^0,  x^1,  x^2, x^3 )=(ct, x, y, z)=(ct, {\bm{x}})}$. We consider a canonical ensemble of massive particles in $O$ with a mean square absolute speed $\langle v^2 \rangle$ and a mean square displacement $\langle d^2 \rangle$, where the mean square displacement vector is defined as
\begin{align}
\langle  {\bm{d}}^2 \rangle \equiv  \left(\langle  d^2 \rangle^x, \langle  d^2 \rangle^y, \langle  d^2 \rangle^z\right) = \left(\langle  x(t)- x(t_0)|^2 \rangle, \langle |y(t)- y(t_0)|^2 \rangle, \langle |z(t)- z(t_0)|^2 \rangle \right)
\end{align}
such that $\langle  d^2 \rangle \equiv  \langle  d^2 \rangle^x + \langle  d^2 \rangle^y + \langle  d^2 \rangle^z$. In the interest of a Lorentz-covariant description of observables, we now parametrize each spacetime event with respect to the proper time $\tau$, such that 
\begin{align}\label{eq: Sigma}
x^\mu \quad \mapsto  \quad x^\mu(\tau) =\left(ct(\tau), x(\tau), y(\tau), z(\tau)\right).
\end{align}
and generalize the mean square displacement to a four-vector, referred to as the mean square spacetime event $\langle  d^2 \rangle^ \mu $, such that
\begin{align}
\langle  d^2 \rangle^\mu \equiv  \left( c^2\langle  t^2 \rangle, \langle  {\bm{d}}^2 \rangle \right),
\end{align}
where $\langle t^2 \rangle$ is the mean square temporal displacement and is defined analogously to its spatial counterparts, i.e., $\langle t^2 \rangle \equiv \langle|t(\tau)-t(\tau_0)|^2 \rangle$. The ensemble is defined by the hypersurface $\mathcal{S}$ of constant proper time in $O$, such that 
\begin{align}
\mathcal{S}  &: \quad  c^2\langle t^2 \rangle - \langle  d^2 \rangle = c^2(\Delta\tau)^2, \label{eq: spacetime surface}
\end{align}
where $\Delta \tau \equiv \tau-\tau_0$. By definition, the ratio of absolute time to proper time is the Lorentz factor $\gamma$. Given in terms of the mean square temporal displacement, we define the mean square Lorentz factor as
\begin{align}
	\langle \gamma^2 \rangle \equiv  \frac{\langle  t^2 \rangle }{(\Delta\tau)^2} = \frac{1}{1-\langle \beta^2_v \rangle},
\end{align}
where $\langle \beta^2_v \rangle \equiv \langle v^2 \rangle/ c^2$ and $\langle v^2 \rangle= \langle  d^2 \rangle / \langle t^2 \rangle$ is the Lorentz-covariant form of the mean square speed, where $\langle  v^2 \rangle < c^2$ (See Appxs. \ref{appendix A} and \ref{appendix B}). The ensemble's relativistic mean square four-velocity can be written as
\begin{align} \label{eq: covariant msv}
	\langle u^2 \rangle ^\mu \equiv  \frac{\langle d^2\rangle ^\mu}{(\Delta\tau)^2}= \left(c^2\langle \gamma^2 \rangle, \langle  {\bm{u}}^2 \rangle \right),
\end{align}
where $\langle u^2 \rangle=\langle d^2\rangle / (\Delta \tau)^2$ is the relativistic mean square speed. Parametrizing the mean square spacetime event $\langle d^2 \rangle ^\mu $ by substituting $\langle  \beta_v^2 \rangle= \langle  d^2 \rangle/\langle t^2 \rangle$ into Eq. (\ref{eq: spacetime surface}) yields
\begin{align}\label{eq: coupled parametrization}
c^2\langle t^2 \rangle - \langle d^2 \rangle  = c^2(\Delta\tau)^2,  \quad \quad \langle \beta_v^2 \rangle = \frac{\langle  d^2 \rangle}{c^2\langle t^2 \rangle }  \quad \quad : \quad \quad 
 c^2\langle t^2 \rangle  = \frac{c^2(\Delta\tau)^2}{1-\langle \beta^2_v \rangle}, \quad \quad \langle  d^2 \rangle = \frac{c^2(\Delta\tau)^2 \langle  \beta_v^2 \rangle}{1-\langle \beta^2_v \rangle}.
\end{align}
In an analogous manner, we can define the mean square quantities of the ensemble's current-density dispersion relation, where $\langle \rho^2 \rangle \equiv \langle | \rho(\tau) - \rho(\tau_0)|^2 \rangle$ is the mean square fluctuation in the observed particle density and $\langle j^2 \rangle \equiv \langle | j(\tau) - j(\tau_0)|^2 \rangle$ is the mean square current density along the radial axis, such that
\begin{equation}\label{eq: rel 1}
\begin{aligned}
c^2\langle \rho^2 \rangle  - \langle j^2 \rangle = c^2\rho_0^2, \quad \quad  \quad \langle \beta^2_v \rangle = \frac{\langle j^2 \rangle}{c^2\langle \rho^2 \rangle}  \quad \quad : \quad \quad
c^2\langle \rho^2 \rangle = \frac{c^2\rho_0^2}{1-\langle \beta^2_v \rangle},  \quad \quad  \quad \langle j^2 \rangle = \frac{c^2\rho_0^2  \langle \beta^2_v \rangle}{1-\langle \beta^2_v \rangle}.
\end{aligned}
\end{equation}
Here we recognize a serendipitous correspondence between the above equations and the mean value relations of the NBD. That is, by parametrizing $\langle \mathcal{N} \rangle - \langle n \rangle = k$ with $\langle p \rangle=\langle n \rangle/\langle \mathcal{N} \rangle$, we can observe that
\begin{equation}\label{eq: stat 1}
\begin{aligned}
\langle \mathcal{N} \rangle - \langle n \rangle = k,  \quad \quad  \langle p \rangle  = \frac{\langle n \rangle}{\langle \mathcal{N} \rangle}  \quad\quad \quad : \quad\quad \quad
\langle \mathcal{N} \rangle = \frac{k}{1-\langle p \rangle},  \quad \quad   \langle n \rangle = \frac{k\langle p \rangle}{1 - \langle p \rangle}.
\end{aligned}
\end{equation}
As the dynamical equations of (\ref{eq: rel 1}) and statistical equations of (\ref{eq: stat 1}) differ only by the labels of their arguments---i.e., they are isomorphic---we can establish the following bijective map:
\begin{align}\label{eq: correspondence}
c^2\rho_0^2 \mapsto k, \quad \quad \langle \beta^2_v \rangle \mapsto \langle p \rangle, \quad \quad c^2\langle\rho^2 \rangle  \mapsto \langle \mathcal{N} \rangle, \quad \quad \langle j^2 \rangle \mapsto \langle n \rangle.
\end{align}
As we later confirm, this surprising correspondence is more than a coincidence, as we recover precisely the same result from quantum mechanics. This identifies the square current density $j^2$ as the random variable to be observed, and the probability of its observation is given by the relative mean square speed. The rest density $\rho_0^2$, as related to the dispersion parameter $k$, serves as the alternative to an observation of the square current density. In this interpretation, for any given interval of time, one either measures the particle density itself or its relativistic rate of change per unit time. A similar correspondence to Eq. (\ref{eq: correspondence}) could likewise be drawn between the mean NBD parameters and the ensemble's mean square energy/momentum (See Eq. (\ref{eq: sp NBD simple})), or the particles' mean square temporal/spatial displacements via Eq. (\ref{eq: coupled parametrization}), or any physical quantities inherent to such a hyperbolic equation; however, the fundamental relationship underlying each correspondence is actually dimensionless. Therefore, more generally, we define a unitless relation by dividing $c^2\langle t^2 \rangle - \langle d^2 \rangle = c^2(\Delta\tau)^2$ by the square of the Planck length, $\ell_p^2\equiv\hbar G /c^3$, to yield the equation $\langle \zeta^2 \rangle - \langle \xi^2 \rangle = \varsigma^2$, where
\begin{align}\label{eq: unitless}
\varsigma^2 \equiv \frac{c^2 (\Delta\tau)^2}{\ell_p^2}, \quad \quad \langle \zeta^2 \rangle\equiv \frac{c^2\langle t^2 \rangle}{\ell_p^2}, \quad \quad \langle \xi^2 \rangle\equiv \frac{\langle d^2 \rangle}{\ell_p^2}    \quad \quad : \quad \quad \langle \beta^2_v \rangle=\frac{\langle \xi^2 \rangle}{\langle \zeta^2 \rangle},
\end{align}
and therefore,
\begin{align}\label{eq: NBD correspondence}
\varsigma^2 = k, \quad \quad \langle \beta^2_v \rangle = \langle p \rangle, \quad \quad \langle \zeta^2 \rangle = \langle \mathcal{N} \rangle, \quad \quad \langle \xi^2 \rangle = \langle n \rangle.
\end{align}
It should be emphasized that the factor of $\ell_p^{-2}$ in Eq. (\ref{eq: unitless}) was not an ad hoc choice but in fact a physically-significant one, which will become apparent in the context of the path-integral derivation presented in the following section (See the comments following Eq. (\ref{eq: gamma kernel}) for more details).

The latter three identities of Eq. (\ref{eq: NBD correspondence}) equate the first and second central statistical moments of their respective parameters. In terms of physical quantities, we refer to $\varsigma$ as the \textit{protodensity} of the system at rest, $\zeta$ as the relativistic or \textit{proper protodensity} of the system in motion, and $\xi$ as the system's relativistic \textit{protocurrent}---in other words, the vacuum precursors of physical densities/currents in Minkowski space. In terms of these parameters, we can write the first expression of Eq. (\ref{eq: NBD}) as
\begin{align}\label{eq: NBD radial case}
\Pr\left(\xi^2\right) =  \frac{\Gamma{(\xi^2+\varsigma^2)}}{\xi^2!\Gamma{(\varsigma^2)}} \left(1-\langle \beta^2_v \rangle\right)^{\varsigma^2} (\langle \beta^2_v \rangle)^{\xi^2}.
\end{align}
Alternatively, when $\varsigma$ is given units of a mass density $m$, such that the NBD evaluates the probability of observing the ensemble's instantaneous square momentum $\text{p}^2$, these results produce a particularly symmetric expression in terms of the energy $\text {E}$. Setting $c=1$, we find
\begin{align}\label{eq: sp NBD simple} 
 \Pr(\text{p}^2) &=  \frac{1}{\tilde{\text{p}}{}^2!}\frac{\Gamma\big({\tilde{\text{E}}}{}^2\big)}{\Gamma\big( \tilde m{}^2 \big)} \left(\frac{\tilde m{}^2 }{\langle \tilde{\text{E}}{}^2 \rangle} \right)^{\tilde m{}^2} \left(\frac{\langle\tilde{ \text{p}}{}^2 \rangle}{\langle \tilde{\text{E}}{}^2 \rangle} \right)^{\tilde{\text{p}}{}^2},
\end{align}
where $\langle v^2 \rangle$ has been parametrized in terms of $\langle \text{p}^2\rangle$ and $\langle \text{E}^2\rangle$, and $\tilde \cdot$ denotes multiplication by an appropriate factor to make each expression unitless. Due to the factorial expression, it is required that the square momentum equals a positive integer and is therefore quantized.\\

\paragraph*{Generalized correspondence in 3+1 dimensions---} \label{sec: Generalized Correspondence in 4-Space} We now generalize the above radial correspondence to describe the full spacetime metric. The ensemble's hypersurface parametrization becomes
\begin{align}
\mathcal{S} \quad : \quad c^2\langle t^2 \rangle - \langle d^2 \rangle^x - \langle d^2 \rangle^y - \langle d^2 \rangle^z = c^2(\Delta\tau)^2,
\end{align}
and the dimensionless mean square representation is the quadratic form
\begin{align}
\langle \zeta^2 \rangle - \langle \xi^2 \rangle^x - \langle \xi^2 \rangle^y - \langle \xi^2 \rangle^z = \varsigma^2,
\end{align}
which, in terms of statistical parameters, is to be expressed as
\begin{align}
\langle \mathcal{N} \rangle  - \langle n \rangle^x - \langle n \rangle^y - \langle n \rangle^z = k.
\end{align}

The intention is to define a unique NBD along each axis of space while remaining consistent with the radial distribution derived above. To this end, we define the mean statistical parameters as the sum of their axial components, such that
\begin{align}
\langle p \rangle = \sum_{i=1}^3 \langle p \rangle^i, \quad  \quad \langle n \rangle = \sum_{i=1}^3 \langle n \rangle^i, \quad   \quad \langle \mathcal{N} \rangle = \sum_{i=1}^3 \langle \mathcal{N} \rangle^i, 
\end{align}
and where the value of $k$ is uniform among the axial and radial distributions. Given the hypersurface constraint of the ensemble, one can define a mean square velocity vector $\langle \bm{v}^2 \rangle\equiv  (\langle  v^2 \rangle^x, \langle  v^2 \rangle^y, \langle  v^2 \rangle^z )$, such that $\langle  v^2 \rangle^i =  \langle  d^2 \rangle^i / \langle t^2 \rangle$ (See Appx. \ref {appendix B}). As the canonical ensemble is in thermodynamic equilibrium, we choose the inertial reference frame where the ensemble's mean square velocity along each axis is uniform, i.e., $\langle v^2\rangle^x = \langle v^2\rangle^y = \langle v^2\rangle^z = \langle v^2\rangle/3$. In order to define an axial NBD, we must reformulate the relativistic velocity along each axis into a form analogous to the radial expression. That is, a component of the relativistic velocity must be expressed in terms of a single parameter as it is in the radial case. Writing the $i$th component of the mean square relativistic velocity as the dimensionless factor $\langle \beta_u^2\rangle^i = \langle u^2\rangle^i/c^2$, we have
\begin{align}
\langle \beta_u^2\rangle^i = \frac{\langle \beta_v^2\rangle/3}{1-\langle \beta_v^2\rangle} = \frac{\langle \beta_v^2\rangle^{\prime}}{1-\langle \beta_v^2\rangle^{\prime}} = \frac{\langle p \rangle'}{1-\langle p \rangle'},
\end{align}
where $\langle p \rangle'$ is the mean chance along each axis. It follows that the effective mean square axial velocity $\langle \beta_v^2\rangle^{\prime}$ is
\begin{align}
\langle \beta_v^2\rangle^{\prime} \equiv \frac{\langle \beta_v^2 \rangle}{3-2\langle \beta_v^2 \rangle} \quad \Leftrightarrow \quad \langle p \rangle' \equiv \frac{\langle p \rangle}{3-2\langle p \rangle}.
\end{align}
We can now express the components of $\langle \zeta^2 \rangle$ and $\langle \xi^2 \rangle$, and their statistical counterparts $\langle \mathcal{N} \rangle$ and $\langle n \rangle$ as
\begin{align}
\langle \zeta^2 \rangle^i = \frac{\varsigma^2}{1-\langle \beta_v^2\rangle^{\prime}}  \quad \Leftrightarrow \quad \langle \mathcal{N} \rangle^i = \frac{k}{1-\langle p \rangle'}, \quad\quad\quad
\langle \xi^2 \rangle^i = \frac{\varsigma^2 \langle \beta_v^2\rangle^{\prime}}{1-\langle \beta_v^2\rangle^{\prime}}  \quad \Leftrightarrow \quad \langle n \rangle^i = \frac{k\langle p \rangle'}{1-\langle p \rangle'}.
\end{align}

Therefore, the mean number of trials $\langle \mathcal{N} \rangle$ and the mean number of successes along each axis $\langle \bm{n} \rangle$ can be collected into a four-vector $\langle n \rangle^\mu$, which is equivalent to the mean square protocurrent four-vector $\langle \xi^2 \rangle^\mu$:
\begin{align}
\langle \xi^2 \rangle^\mu \equiv \left(\langle \zeta^2 \rangle, \langle \bm{\xi}^2 \rangle \right) \quad \Leftrightarrow \quad \langle n \rangle^\mu \equiv (\langle \mathcal{N} \rangle, \langle \bm{n} \rangle).
\end{align}

We can now define the distributions associated with the spatial components of a random measurement $n^\mu = ({\mathcal{N}}, \bm{n})$, which likewise describe the component probabilities associated with the measurement of $(\xi^2)^\mu=(\zeta^2, \bm{\xi}^2)$, such that
\begin{align}
X^i \sim \text{NB}(k, \langle p \rangle' ) \quad&: \quad f_{\text{\tiny{NB}}}(n; k, \langle p \rangle' ) \equiv  \Pr(X^i= n) = \frac{\Gamma{(n+k)}}{k!\Gamma{(k)}}(1-\langle p \rangle')^{k}(\langle p \rangle')^{n},\\
X^i\sim \text{NB}\left(\varsigma^2 ,\langle \beta^2_v \rangle'\right) \quad &: \quad  f_{\text{\tiny{NB}}}(\xi^2; \varsigma^2,\langle \beta^2_v \rangle')  \equiv \Pr\left(X^i=\xi^2\right) =  \frac{\Gamma{(\xi^2+\varsigma^2)}}{\xi^2!\Gamma{(\varsigma^2)}} \left(1-\langle \beta^2_v \rangle'\right)^{\varsigma^2} (\langle \beta^2_v \rangle')^{\xi^2}.\label{eq: NBD 3 space}
\end{align}

\subsection*{Derivation via the Path Integral} \label{sec: Path integral approach}

The correspondence between a statistical event of the NBD and a relativistic event in spacetime was derived in terms of considering the free propagation of a particle system in Minkowski space. Here we prove that such a relationship is consistent with quantum theory by deriving the same result from calculating the kernel, or propagator, of a relativistic system via the path integral. 

One can arrive at this conclusion by considering the spacetime generalization of a typical path integral \cite{struckmeier2009extended},
\begin{align}
\psi(x^\mu_b) = \int_{\mathbbm{R}^4} \text{d}^4x_a  \ \psi(x^\mu_a) K(b| a), \label{eq: path integral}
\end{align}
where $K(b| a)$ is the kernel. As formulated in Feynman's path integral approach to quantum mechanics, the kernel expresses the probability amplitude that a particle transitions from some initial coordinate $x^\mu_a \equiv  (t_a, \bm{x}_a)$ to some final coordinate $x^\mu_b \equiv  (t_b, \bm{x}_b)$ \cite{feynman2010quantum}. In a Lorentz-invariant formulation, the kernel is integrated over both time and space \cite{struckmeier2009extended}, and can be expressed, in general, as
\begin{align}
	K(b| a) = \int_{a}^{b} \mathscr{D}^4 x \ 
\exp\left[ \frac{i}{\hbar} S(x^\mu) \right], \end{align} 
where $S(x^\mu)$ is the action. For our purposes, we will consider the case of a free relativistic particle. The relativistic action \cite{landau2013classical} that is often cited is of the form
\begin{align}
	S = \int _{t_a}^{t_b} \text{d} t \ L\left(t, \bm{x}, \frac{\text{d} \bm{x}}{\text{d} t}\right) = -mc^2 \int_{t_a}^{t_b} \text{d} t \ \sqrt{1-\frac{1}{c^2}\left(\frac{\text{d} \bm{x}}{\text{d} t}\right)^2}. 
\end{align}
While the above expression describes the correct relativistic action, it is not Lorentz-covariant in this form. In order to put both space and time on an equal footing, we parametrize each spacetime coordinate in terms of the proper time $\tau$, such that
\begin{align}
	x^\mu \quad \mapsto \quad  x^\mu(\tau) = \big(ct(\tau), \bm{x}(\tau)\big),
\end{align}
and then use the Struckmeier \textit{extended} Lagrangian $L_e$ \cite{struckmeier2009extended}---here with metric signature $(-,+,+,+)$---defined as
\begin{align}
	L_e\left(x^\mu, \frac{\text{d} x^\mu}{\text{d}\tau}\right) \equiv \frac{m}{2}\left( \frac{\text{d} x_\mu}{\text{d}\tau} \frac{\text{d} x^\mu}{\text{d} \tau} - c^2 \right),\label{eq: extended lagrangian}
\end{align}
where the ``summation convention" is implied. The extended Lagrangian $L_e$ is related to the conventional Lagrangian $L$ by
\begin{align}
L\left(t, \bm{x}, \frac{\text{d} \bm{x}}{\text{d} t}\right)\frac{\text{d} \tau}{\text{d} t}  = L_e\left(x^\mu, \frac{\text{d} x^\mu}{\text{d} \tau}\right), \label{eq: lagrangians}
\end{align}
where $\text{d} \tau/\text{d}t = \gamma^{-1}$. Note that as the extended Lagrangian is not homogeneous to first order in its velocities as required, it must satisfy the constraint equation 
\begin{align}
	\frac{\text{d} x_\mu}{\text{d} \tau} \frac{\text{d} x^\mu}{\text{d}\tau} = -c^2, \label{eq: constaint}  %\left(\frac{\text{d} t}{\text{d} s}\right)^2 - \frac{1}{c^2}\left(\frac{\text{d}\bm{x}}{\text{d} s}\right)^2 = 1, \label{eq: constaint} 
\end{align}
which amounts to parametrizing the velocity along a hypersurface, such that the four-velocity has a constant length---consistent with the velocity parametrization featured in this work. By substituting Eq. (\ref{eq: constaint}) into (\ref{eq: extended lagrangian}), one can check that the relation (\ref{eq: lagrangians}) between Lagrangians is indeed satisfied. Therefore, the extended Lagrangian captures the \textit{same} dynamics as the conventional one (See \cite{struckmeier2009extended} for a rigorous treatment).

One can now express a Lorentz-covariant action in terms of the extended Lagrangian, which splits into a sum of independent action functionals, such that
\begin{align}
	S_e[x^\nu(\tau)] = \frac{m}{2} \int_{\tau_a}^{\tau_b} \text{d} \tau \ \left(\frac{\text{d} x_\mu}{\text{d} \tau} \frac{ \text{d} x^\mu}{\text{d} \tau} -c^2 \right) = \sum_\mu S[x^\mu(\tau)].
\end{align}

In an analogous manner to the conventional path integral approach of discretizing the action into a sum of intervals in \textit{absolute} time, the extended path integral can be expressed by discretizing its extended action integral into a sum over finite intervals of \textit{proper} time $\Delta \tau_\epsilon$. 

The kernel of the relativistic free particle can be expressed, up to a phase factor, by the product of the kernels along each component of spacetime, such that
\begin{align}\label{eq: kernel sup}
\text{K}(b| a) = \exp\left[\frac{imc^2(\tau_b-\tau_a)}{2\hbar}\right] \prod_{\mu=0}^3 K(x^\mu_b|x^\mu_a). 
\end{align}
Each \textit{component kernel} can be expressed by the usual path integral of a free particle but now in terms of $\tau$. We will first evaluate the $x$-component kernel $K(x_b| x_a)$, which is equivalent to $K(y_b| y_a)$ and $K(z_b| z_a)$, and then subsequently evaluate the temporal component kernel $K(t_b| t_a)$. The kernel $K(x_b| x_a)$ is
\begin{align}
K(x_b| x_a) &= \lim_{\Delta \tau_{\epsilon} \to 0} \left( \frac{m}{2\pi i\hbar  \Delta \tau_{\epsilon}} \right)^{N/2} \int \dots \int \text{d} ^{N-1} \! x \ \exp\left[\frac{im}{2\hbar} \sum_{j=1}^N \frac{(x_j - x_{j-1})^2}{\Delta \tau_{\epsilon}} \right], \label{eq: coordinate path integral kernel}
\end{align}
where the $N-1$ integrals are evaluated with respect to $\text{d} ^{N-1} \! x \equiv \text{d} x_{N-1} \dots \text{d} x_j \dots \text{d} x_1$ over $\mathbbm{R}^{N-1}$, with $x_a\equiv x_0$ and $x_b\equiv x_N$, and where $N \Delta \tau_{\epsilon} = \tau_b-\tau_a$ is the proper time interval of propagation. 

The solution of the above integral is well known and can be solved analytically in various ways \cite{feynman2010quantum} \cite{shankar2012principles}, yielding
\begin{align}
K(x_b| x_a)  &= \left( \frac{m}{2\pi i \hbar  (\tau_b-\tau_a)} \right)^{1/2} \exp\left[\frac{im(x_b - x_{a})^2}{2\hbar(\tau_b-\tau_a)} \right]. \label{eq: kernel solution}
\end{align}
However, instead of evaluating the kernel in this manner, here an alternative expansion of Eq. (\ref{eq: coordinate path integral kernel}) is considered as follows.\\

We first employ a common technique by evaluating the Euclidean path integral via a Wick rotation \cite{wick1954properties}, such that $\tau \equiv - i  \uptau$. The resulting expression is equivalent to a diffusion relation. Through the Wick rotation, one can interpret the change in time as the inverse temperature, such that $\Delta \uptau_\epsilon/ \hbar=1/(k_B T)$, where $k_B$ is the Boltzmann constant \cite{zee2010quantum}. 

With the above changes, the kernel of Eq. (\ref{eq: coordinate path integral kernel}) becomes 
\begin{align}
K(x_b| x_a) &= \lim_{\Delta \uptau_{\epsilon} \to 0} \left( \frac{m }{2\pi k_B T(\Delta\uptau_\epsilon)^2} \right)^{ N/2} \int_{\mathbb{R}^{N-1}} \text{d} ^{N-1} \! x \ \exp\left[-\frac{m}{2 k_BT} \sum_{j=1}^{ N} \left(\frac{\Delta  x_{j}}{\Delta\uptau_\epsilon}\right)^2 \right],
\end{align}
where $(\Delta x_{j})^2 \equiv (x_{ j} -  x_{j-1})^2$. The free particle kernel is normalized with respect to its final location, such that
\begin{align}
\int_{-\infty}^{\infty} \text{d}  x_b \ K(x_b| x_a) = 1.
\end{align}
This is equivalent to extending the $N-1$ integrals that define a kernel to a total of $N$. Applying this to the free particle integral and making the substitution ${\text{d} x_{j} \to \text{d}(\Delta x_{j} / \Delta\uptau_\epsilon)}$, we have
\begin{align}
\int_{-\infty}^{\infty} \text{d} x_b \  K(x_b| x_a) &= \lim_{\Delta\uptau_{\epsilon} \to 0} \left( \frac{m 
}{2\pi k_B T} \right)^{N/2} \int_{\mathbb{R}^{N}} \text{d} ^{N} \! \left(\frac{\Delta x}{\Delta\uptau_\epsilon}\right) \ \exp\left[-\frac{m}{2 k_B T} \sum_{j=1}^{N} \left(\frac{\Delta x_{j}}{\Delta\uptau_\epsilon}\right)^2 \right].
\end{align}
The above expression is equivalent to the integral over the Maxwell-Boltzmann distribution in $N$ dimensions but physically describes particle propagation purely along the $x$-axis. Mathematically, each of the $N$ subintervals of the total trajectory effectively behaves as an independent dimension of propagation.\\

We will now drop the limit and consider propagation on a lattice as an approximation to a smooth trajectory, where we replace $\Delta \uptau_\epsilon$ with the Planck time $\Delta \uptau_p=\sqrt{\hbar G / c^5}$, i.e., the smallest time interval consistent with quantum theory.\\

Using Fubini’s theorem, while exploiting the hyperspherical symmetry of the integrand, one can reduce the $N$ Gaussian integrals to a single integral of a ``radial-like" variable $\Delta x_r / \Delta\uptau$, where $(\Delta x_r)^2 \equiv \sum_{j=1}^N \tfrac{N}{2}(\Delta x_j)^2$ and ${(\Delta \uptau)^2 \equiv \tfrac{N}{2}(\Delta \uptau_p)^2}$, such that the expression becomes
\begin{align}
\left( \frac{m }{2\pi k_B T} \right)^{N/2} \frac{2\pi^{N/2}}{\Gamma{\left(\frac{N}{2}\right)}} \int_{0}^{\infty} \text{d}\left(\frac{\Delta x_r}{\Delta \uptau}\right) \ \left(\frac{\Delta x_r}{\Delta\uptau}\right) ^{N-1}\exp\left[-\frac{m }{2 k_B T}  \left(\frac{\Delta x_r}{\Delta \uptau}\right)^2 \right].
\end{align}
We can express the above in terms of the mean square displacement of $x$, where $\langle d^2 \rangle^x = (\Delta x_r)^2 / N$; however, it should be noted that this is not a static average but is dependent upon the integration variable. Furthermore, by generalizing the typical identification of the mean square absolute speed with the quantity $k_B T/m$ of a particle ensemble in one dimension \cite{callen1998thermodynamics}, we now identify $\langle u^2 \rangle^x =  k_B T/m$ as the mean square \textit{relativistic} speed along the $x$-axis. In addition to applying these identities, we replace $N/2$ by the notation $\varsigma^2 \equiv  N / 2$, introduce relevant factors of $c$, and further simplify the expression by another change-of-variable substitution, integrating now with respect to $\varsigma^2 \langle d^2 \rangle^x/ (c\Delta\uptau)^2 $, such that
\begin{align}
\int_{0}^{\infty} \text{d} \left( \frac{\varsigma^2\langle d^2 \rangle^x }{c^2(\Delta\uptau)^2}\right)\ \left(\frac{c^2}{\langle u^2 \rangle^x}\right)^{\varsigma^2} \frac{\left( \frac{\varsigma^2\langle d^2 \rangle^x}{c^2(\Delta\uptau)^2}\right)^{\varsigma^2-1}}{\Gamma{(\varsigma^2)}} \exp\left[-\frac{c^2}{\langle u^2 \rangle^x}\frac{\varsigma^2\langle d^2 \rangle^x }{c^2(\Delta\uptau)^2}\right].
\end{align}

The resulting expression is precisely the integral over the gamma distribution:
\begin{align}\label{eq: gamma kernel}
\int_{-\infty}^{\infty} \text{d} x_b \ K(x_b| x_a) = \int_{0}^{\infty} \text{d} \langle \hat \xi^2 \rangle \ f_{\text{\tiny{G}}} (\langle \hat \xi^2 \rangle; \varsigma^2, \langle \hat \beta^2_u \rangle),
\end{align}
where we make the identification $\langle \hat \xi^2 \rangle = \varsigma^2 \langle \hat d^2 \rangle / (c\Delta\uptau)^2$, and where we have replaced the $x$-component superscript with $\hat \cdot$ to reduce clutter. Despite the complicated integration variable, it should be noted that the only non-constant parameter of the integral is $\langle \hat d^2 \rangle$.

As suggested by the notation, the quantity $\varsigma^2=N/2$ that appears as the shape parameter in the gamma distribution is in fact the same rest protodensity parameter we have identified with the dispersion parameter $k$ of the NBD, which gives the quantity of $(\Delta \uptau_p)^2$ intervals in the radially-symmetric expression of the Euclidean path integral. It follows that the random variable $\langle \hat \xi^2 \rangle$ of the above distribution can be identified with the protocurrent of the particle ensemble along the $x$-axis, as defined in the previous section, as $\langle \hat \xi^2 \rangle = \varsigma^2 \langle \hat d^2 \rangle / (c\Delta\uptau)^2 = \langle \hat d^2 \rangle/\ell_p^2$, which is consistent with the definition in Eq. (\ref{eq: unitless}). It is easy to check that $\langle \hat \zeta^2 \rangle$ is also consistently defined in a similar manner.

Written now in the more concise notation, the integrand of the kernel expresses the distribution
\begin{align}
f_{\text{\tiny{G}}} (\langle \hat \xi^2 \rangle; \varsigma^2,\langle \hat \beta^2_u \rangle) = \frac{\langle \hat \xi^2 \rangle^{\varsigma^2-1}}{\langle \hat \beta^2_u \rangle^{\varsigma^2}\Gamma{(\varsigma^2)}} e^{-\langle \hat \xi^2 \rangle / \langle \hat \beta^2_u \rangle}.
\end{align}
There is a discrete version of Eq. (\ref{eq: gamma kernel}) given in terms of the instantaneous $\xi$-vector, and it comes from this kernel for free. Before integrating, one must simply multiply the equation by a factor of 1, expressed in the form of an exponential times its inverse and expanded in its Maclaurin series, i.e., the sum over the Poisson distribution, such that
\begin{align}
\int_0^{\infty} \text{d}\langle \hat \xi^2 \rangle \ f_{\text{\tiny{G}}} (\langle \hat \xi^2 \rangle; \varsigma^2, \langle \hat \beta^2_u \rangle)&= \int_0^{\infty} \text{d}\langle \hat \xi^2 \rangle \ \frac{\langle \hat \xi^2 \rangle^{\varsigma^2-1}}{\langle \hat \beta^2_u \rangle^{\varsigma^2}\Gamma{(\varsigma^2)}} e^{-\langle \hat \xi^2 \rangle/\langle \hat \beta^2_u \rangle}\\
&= \int_0^{\infty} \text{d}\langle \hat \xi^2 \rangle \ e^{\langle \hat \xi^2 \rangle} e^{-\langle \hat \xi^2 \rangle} \frac{\langle \hat \xi^2 \rangle^{\varsigma^2-1}}{\langle \hat \beta^2_u \rangle^{\varsigma^2}\Gamma{(\varsigma^2)}} e^{-\langle \hat \xi^2 \rangle / \langle \hat \beta^2_u \rangle}\\
&= \int_0^{\infty} \text{d}\langle \hat \xi^2 \rangle \ \sum_{{n}=0}^{\infty} \frac{\langle \hat \xi^2\rangle^{{n}}}{{n}!} e^{-\langle \hat \xi^2 \rangle} \frac{\langle \hat \xi^2 \rangle^{\varsigma^2-1}}{\langle \hat \beta^2_u \rangle^{\varsigma^2}\Gamma{(\varsigma^2)}} e^{-\langle \hat \xi^2 \rangle  / \langle \hat \beta^2_u \rangle}\\
&= \sum_{{n}=0}^{\infty} \frac{\langle \hat \beta^2_u \rangle^{-\varsigma^2}}{{n}!\Gamma{(\varsigma^2)}} \int_0^{\infty} \text{d}\langle \hat \xi^2 \rangle \ \langle \hat \xi^2 \rangle^{{n}+\varsigma^2-1} e^{-\langle \hat \xi^2 \rangle(1 + 1 / \langle \hat \beta^2_u \rangle)}.
\end{align}
We now write the mean square relativistic speed in terms of the effective mean square absolute speed as measured in the given inertial reference frame, such that $\langle \hat \beta^2_u \rangle =  \langle \hat \beta^2_v \rangle'/(1- \langle \hat \beta^2_v \rangle')$. The kernel can now be expressed as
\begin{align}
\int_0^{\infty} \text{d}\langle \hat \xi^2 \rangle \ f_{\text{\tiny{G}}} (\langle \hat \xi^2 \rangle; \varsigma^2,   \langle \hat \beta^2_u\rangle )
&= \sum_{{n}=0}^{\infty} \frac{(1-  \langle \hat \beta^2_v \rangle')^{ \varsigma^2}}{ \langle \hat \beta^2_v \rangle^{\prime \varsigma^2} {n}! \Gamma{(\varsigma^2)}} \int_0^{\infty} \text{d}\langle \hat \xi^2 \rangle \ \langle \hat \xi^2 \rangle^{{n}+\varsigma^2-1} e^{-\langle \hat \xi^2 \rangle/   \langle \hat \beta^2_v \rangle'}\\
&= \sum_{{n}=0}^{\infty} \frac{(1- \langle \hat \beta^2_v \rangle')^{\varsigma^2}}{ \langle\hat \beta^2_v \rangle^{\prime \varsigma^2} {n}! \Gamma{(\varsigma^2)}}   \langle \hat \beta^2_v \rangle^{\prime \varsigma^2+{n}} \Gamma({n}+\varsigma^2)\\
&= \sum_{{n}=0}^{\infty} \frac{\Gamma(n+\varsigma^2)}{{n}! \Gamma{(\varsigma^2)}} (1-  \langle \hat\beta^2_v \rangle')^{\varsigma^2}   \langle\hat \beta^2_v \rangle^{\prime {n}},
\end{align}
which is precisely the sum over the negative binomial distribution. This result was derived from the so-called Poisson-gamma mixture. The Poisson rate parameter that was integrated out of the gamma prior was $\langle \hat \xi^2 \rangle$, such that we can identify the random variable of the posterior predictive NB distribution as $\hat\xi^2=n$. Therefore, we recover exactly the component distribution derived in Eq. (\ref{eq: NBD 3 space}):
\begin{align}
f_{\text{\tiny{NB}}} (\hat \xi^2; \varsigma^2,   \langle \hat \beta^2_v \rangle) = \frac{\Gamma({\hat\xi^2}+\varsigma^2)}{{\hat\xi^2}! \Gamma{(\varsigma^2)}} \left(1- \langle  \beta^2_v \rangle'\right)^{\varsigma^2}   \langle \hat\beta^2_v \rangle^{\prime {\hat\xi^2}}.\label{eq: propagator relations}
\end{align}
In summary, we have found that
\begin{align}
\int_{-\infty}^{\infty} \text{d} x_b \ K(x_b| x_a) = \int_0^{\infty} \text{d}\langle \hat \xi^2 \rangle \ f_{\text{\tiny{G}}} (\langle \hat \xi^2 \rangle; \varsigma^2, \langle \hat \beta^2_u \rangle )
= \sum_{\hat \xi^2=0}^{\infty} f_{\text{\tiny{NB}}}(\hat \xi^2; \varsigma^2,  \langle \hat\beta^2_v \rangle'),\label{eq: propagator relation}
\end{align}
where $\varsigma^2=N/2$. As previously stated, $K(y_b| y_a)$ and $K(z_b| z_a)$ are identical to $K(x_b| x_a)$, and we are now just left to evaluate the temporal component kernel $K(t_b| t_a)$. The process is nearly the same but with some substitutions:
\begin{align}
\Delta x_r \mapsto -i\Delta t_r, \quad \quad \langle \beta_u^2 \rangle^x \mapsto \langle \beta_u^2 \rangle^t = -\langle \gamma^2 \rangle, \quad \quad \langle d^2 \rangle^x \mapsto -\langle t^2 \rangle, \quad \quad \langle \xi^2 \rangle^x \mapsto \langle \xi^2 \rangle^t  =\frac{-\varsigma^2 \langle t^2 \rangle}{c^2(\Delta\uptau)^2} =  -\langle \zeta^2 \rangle,
\end{align}
where the factor of $-i$ and subsequent negative signs are due to the Wick rotated time axis. Additionally, we can rewrite the Lorentz factor as
\begin{align}
-\langle \gamma^2 \rangle = \frac{1/\langle \beta_v^2 \rangle}{1-1/\langle \beta_v^2 \rangle}.
\end{align}

\begin{figure}[H]
\caption{Double NBD parameters $\langle p \rangle_1$ and $\langle p \rangle_2$ as a function of a maximum pseudorapidity $|\eta|_{max}$. Charged particle multiplicity data and NBD models provided by the ALICE Collaboration \cite{acharya2017charged}, depicting non-single-diffraction events (NSD), inelastic events of at least one charged particle in the $|\eta|<1.0$ range (INEL $>0$), and all inelastic collision events (INEL).}
  \includegraphics[width=0.95\textwidth]{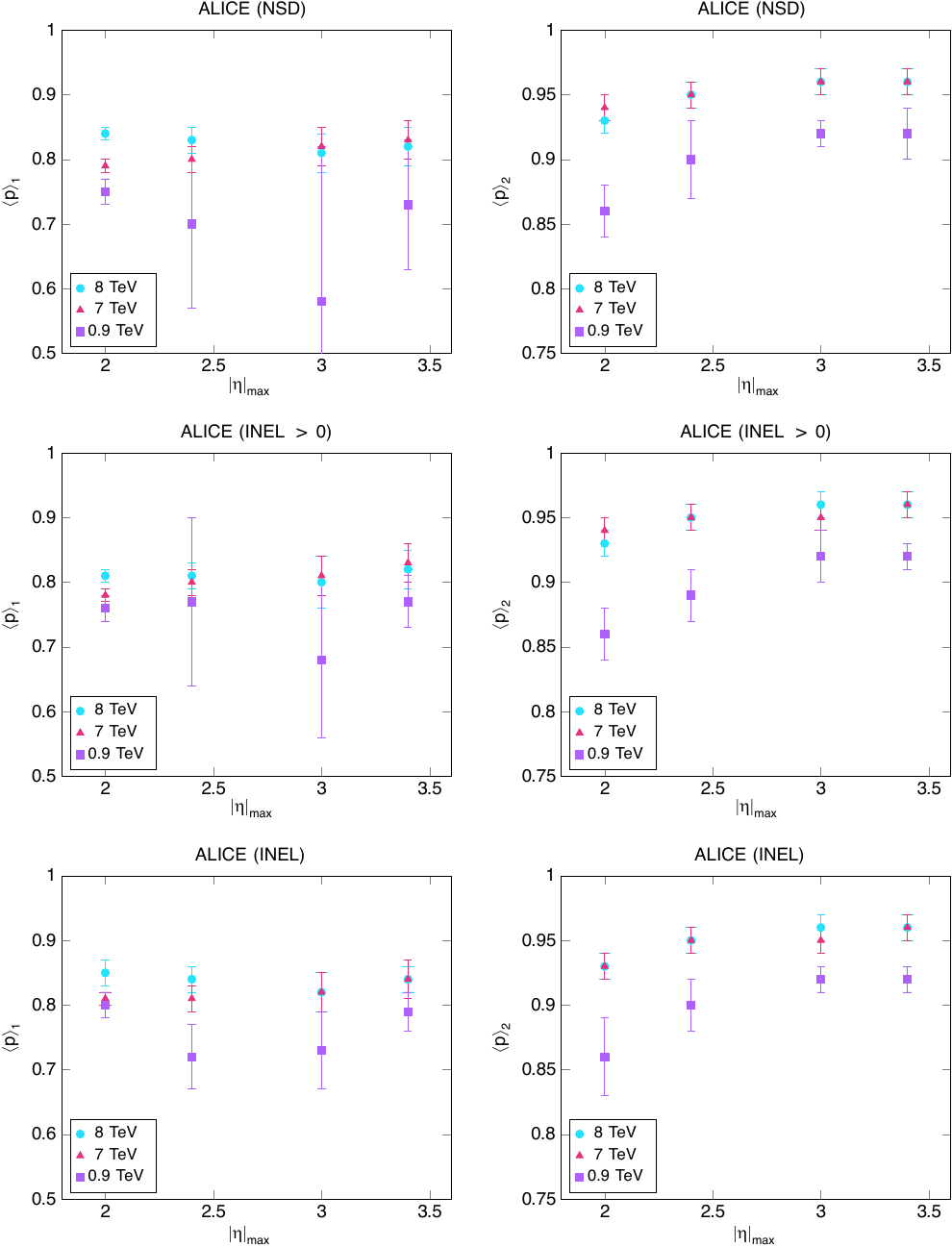}
  \label{fig: p vs eta}
\end{figure}

Therefore, by following all the previous steps, the integral of $K(t_b| t_a)$ over the final time $t_b$ yields
\begin{align}
\int_{-\infty}^{\infty} \text{d} t_b \ K(t_b| t_a) &=  \sum_{{\zeta^2}=0}^{\infty} \frac{\Gamma( -\zeta^2+\varsigma^2)}{(-\zeta^2)! \Gamma{(\varsigma^2)}} \left(1-  \frac{1}{\langle  \beta^{2}_v \rangle}\right)^{\varsigma^2} \left(\frac{1}{\langle  \beta^{2}_v\rangle} \right)^{-\zeta^2}\\
&= \sum_{{\zeta^2}=0}^{\infty} \frac{\Gamma(-\varsigma^2+1)}{(-\varsigma^2+\zeta^2)!\Gamma(-\zeta^2+1)}\left(1- \langle  \beta^{2}_v \rangle\right)^{\varsigma^2} \left(-\langle  \beta^{2}_v\rangle \right)^{\zeta^2-\varsigma^2}\\
&=\sum_{{\xi^2}=0}^{\infty} \frac{\Gamma(\xi^2+\varsigma^2)}{{\xi^2}! \Gamma{(\varsigma^2)}} \left(1- \langle  \beta^{2}_v \rangle\right)^{\varsigma^2}   \langle  \beta^{2}_v \rangle^{\xi^2},
\end{align}
where we have used $\zeta^2 - \xi^2= \varsigma^2$ and the identity (\ref{eq: negative NBD coefficient form}) from Appx. \ref{sec: NBD}, and changed the summation index from $\zeta^2$ to $\xi^2$ by tacitly shifting the value of $\varsigma^2$. Therefore, the temporal component kernel yields the NBD of the radial parameters in perfect agreement with Eq. (\ref{eq: NBD radial case}), such that
\begin{align}
\int_{-\infty}^{\infty} \text{d} t_b \ K(t_b| t_a) &= \sum_{\xi^2=0}^{\infty} f_{\text{\tiny{NB}}}(\xi^2; \varsigma^2,  \langle \beta^2_v \rangle ).
\end{align}
Therefore, the complete spacetime kernel of Eq. (\ref{eq: kernel sup}) can be represented, up to a phase factor, by the product of the NB spatial distributions and NB radial distribution, which are in perfect agreement with the definitions of Eqs. (\ref{eq: NBD radial case}) and (\ref{eq: NBD 3 space}). As the kernel evaluates the probability of a Wiener-L\'evy process of some position-dependent function, it applies equally well to determining the mean statistical fluctuations of the square current density $j^2(x)$, and thus we resolve that these results reproduce the bijective map of Eq. (\ref{eq: correspondence}) both radially and along the cardinal axes when evaluated with this particular dimensionality.

\subsection*{Comparison with ALICE and ATLAS Experiments}

The experimental data of $pp$ collisions produced at the LHC are shown to be consistent with the theoretical model presented in these proceedings, which approximates the set of product particles as relativistic members of a canonical ensemble emanating from the interaction point. In modeling the measurements at the LHC, the protodensity $\varsigma$ is given units of a particle rest density $\rho_0$, as in Eq. (\ref{eq: correspondence}). This establishes the square current density $j^2$ as the random variable of the NBD and is therefore associated with the charged particle multiplicity $n$ measured in a collision experiment. In this model, the parameter $\langle p \rangle$ is equivalent to the mean square speed $\langle \beta_v^2 \rangle$ of the ensemble, and therefore, one should anticipate a positive correlation of $\langle p \rangle$ with both the pseudorapidity of the detected particles and the center-of-mass energy of the collision. Both of these trends are observed in the double NBD fits by the ALICE collaboration \cite{acharya2017charged} (Fig. \ref{fig: p vs eta}, Table \ref{table: ALICE}), where three data sets are compared: non-single-diffraction events (NSD), all inelastic collision events (INEL), and inelastic events of at least one charged particle in the $|\eta|<1.0$ range (INEL $>0$). Particles produced in these collisions travel at speeds reaching a significant percentage of $c$, and therefore we expect to find values of $\langle p \rangle$ reflecting this fact, which are also observed. In particular, we see values close to unity in hard/semi-hard collisions associated with $\langle p \rangle_2$ and to a lesser degree in soft collisions via $\langle p \rangle_1$, as would be expected by the greater pseudorapidity associated with the former as opposed to the latter. This also agrees with the triple NBD model by \cite{Zborovsky2018} (Fig. \ref{fig: NBD MD}) on measurements by the ATLAS group, where the shoulder distribution is believed to represent soft-collisions ($\langle p \rangle < 0.5$) and the tail distributions to capture harder collision events ($\langle p \rangle > 0.9$). In the ALICE data, a positive correlation between $\langle p \rangle_2$ and $|\eta|_{max}$ is evident, especially at 0.9 TeV, but due to the high uncertainty in $\langle p \rangle_1$, the trend between $\langle p \rangle_1$ and $|\eta|_{max}$ in soft collision events is less clear.

The discussed data and existing literature support an inverse relationship between the dispersion parameter $k$ and $\sqrt{s}$ within the range of tested collision energies. Empirically, $k^{-1}$ can be approximated as a linear function in $\ln \sqrt{s}$, but the fundamental reason for this trend remains unknown. According to the present model, $k$ represents the square rest density of the product ensemble, implying that the rest density of the measured ensemble reduces with energy, even as the current increases. This decrease in rest density is compensated by an increase in rapidity. That is, the increased energy is going toward scattering the products at ever-greater speeds while simultaneously suppressing the creation of particle mass. It is important to note, however, that the decrease in the rest density does not imply a decrease in the sample's measured relativistic density with increasing energy. On the contrary, the mean square relativistic density $\langle \rho^2 \rangle$ (corresponding to $\langle \mathcal{N} \rangle=\langle n \rangle + k$) is observed to increase with both $\sqrt{s}$ and $|\eta|_{max}$, as expected. The association of $k$ with the rest density could help improve our understanding of processes at the interaction point, as the relationship of $k$ with $|\eta|$ has been observed to differ among soft and semi-hard collisions. 

While these theoretical results correlate well with experimental data, they nevertheless represent a general model of the physical systems under study, and precise quantitative predictions may require the addition of more nuanced considerations, such as the presence of external potentials and other interactions, in order to accurately describe experimental measurements.

\paragraph*{Concluding Remarks---} We have shown that the mean value relations of the NBD can be derived via the dynamical equations of a free relativistic particle ensemble along a hypersurface of constant proper time, and that the NBD itself is a representation of the quantum mechanical propagator of precisely such a system. These two derivations, despite their disparate assumptions, arrive at exactly the same conclusion: that the expected multiplicity of a relativistic particle ensemble is described by the negative binomial distribution. Moreover, each approach is premised on the most fundamental principles of the theory from which it originates---namely, special relativity and quantum mechanics---thereby demonstrating a robust theoretical argument, from first principles, for the significance of the well-documented NBD multiplicity statistics observed in particle collision experiments. These results have been shown to consistently describe NBD models of experimental data from the ATLAS and ALICE measurements at the LHC. In particular, the parameters attributed to the NBD models have been shown to associate with the variables intrinsic to particle current density, and in such a way that positively correlates with experimental observations. General trends in the rapidity and center-of-mass energy $\sqrt{s}$ of collision products agree with the behavior of the physical parameters related to each, which have been attributed to particular parameters of the NBD. Furthermore, a theoretical motivation for the unknown significance of the inverse relationship of $\sqrt{s}$ and the NBD dispersion parameter $k$ results from this work, which makes an association between $k$ and the ensemble rest density.

In addition to providing a general phenomenological justification for the puzzling occurrence of the NBD in multiplicity measurements, a significant theoretical consequence of these results is that the Minkowski metric equation in its dimensionless form $\langle \zeta^2 \rangle - \langle \xi^2 \rangle^x - \langle \xi^2 \rangle^y - \langle \xi^2 \rangle^z = \varsigma^2$ in fact implies the free particle path integral and vice versa, which establishes an apparent statistical significance to the hyperbolic geometry of Minkowski spacetime. In other words, the flat spacetime metric has been shown to effectively encode the negative binomial statistics of many-body systems, and in direct connection with those as derived via the quantum mechanical path integral formalism. The mean value relation of the NBD expression is shown to manifest from the Minkowski metric equation by dividing the latter by the square of the Planck length, resulting in the dimensionless form referred to above. Alternatively, following an exploitation of the hyperspherical symmetry of the path integral, a simplified expression is achieved, which also yields precisely the negative binomial distribution of interest. Ultimately, this novel connection between the theories of special relativity and quantum mechanics provides a revealing perspective into the coexistence and apparent interdependence of these two theoretical regimes.

Further research is proposed to increase alignment of these theoretical considerations with experimental data by extending the present model to account for additional variables in the systems under study.

\section*{Acknowledgement}
The support toward the realization of this research is acknowledged and attributed to Dr. Rudolf, Hedy, and Avanisha Filz with gratitude and appreciation.

\begin{table}[H]
\caption{NBD parameters of multiplicity data compiled and modeled by the ALICE Collaboration at the LHC \cite{acharya2017charged}. The data represents a double NBD fit over three center-of-mass energies $\sqrt{s}$ (TeV) and five pseudorapidity intervals $|\eta|$. Three data classes are provided, depicting non-single-diffraction events (NSD), inelastic events of at least one charged particle in the $|\eta|<1.0$ range (INEL $>0$), and all inelastic collision events (INEL). To compensate for the different data classes, a parameter $\lambda$ weights each NBD fit, where
$\Pr(n)=\lambda[\alpha f_{NB}(n; \langle n \rangle_1, k_1)+(1-\alpha) f_{NB}(n; \langle n \rangle_2, k_2)].$
Additionally, we display the parameter $\langle p \rangle$ for each NBD, where $\langle p \rangle=\langle n \rangle/(k+\langle n \rangle)$.}\label{table: ALICE}
\begin{subtable}{\textwidth}
\caption*{NSD}\label{table: NSD}
\renewcommand{\arraystretch}{1.4}%
\begin{tabularx}{\textwidth}{l c c*{8}{Y}}
$\sqrt{s}$  & $\eta$ range & $\lambda \ (\pm 0.01)$ & $\alpha$ & $\langle n \rangle_1$ & $k_1$ & $\langle p \rangle_1$ & $\langle n \rangle_2$ & $k_2$ & $\langle p \rangle_2$\\ \hline \\
 $0.9$ & $|\eta|<2.0$& $0.94$ & $0.52 \pm 0.05$ & $10.28 \pm 0.69$ & $3.49 \pm 0.30$ & $0.75 \pm 0.02 $ & $25.08 \pm 1.96$ & $3.96 \pm 0.63$ & $0.86\pm0.02$ \\
 &  $|\eta|<2.4 $ &  $0.95$ &  $ 0.27 \pm 0.12$ &  $ 10.94 \pm  1.14 $ &  $4.76 \pm  2.91 $ &$0.70\pm0.13$ &  $25.35 \pm  4.40$ &  $ 2.80 \pm  0.96$ & $0.90\pm0.03$\\ 
  &  $|\eta|< 3.0 $ & $ 0.94$ &  $0.17 \pm 0.06$ &  $ 13.68 \pm 0.32$ &  $ 9.81 \pm 9.82 $ & $0.58\pm0.24$ &  $29.62 \pm 2.85$ &  $ 2.54 \pm 0.51$ &  $0.92\pm0.01$\\
    &  $|\eta|< 3.4 $ & $0.94$ & $ 0.26 \pm 0.09 $ & $15.80 \pm 0.95$ & $ 5.95 \pm 2.96$ & $0.73\pm0.10$& $ 35.26 \pm 4.35$ & $ 2.98 \pm 0.77$ & $0.92\pm0.02$\\
        &  $-3.4<\eta< +5.0 $ & $0.95$ & $ 0.36 \pm 0.07 $ & $19.27 \pm 1.45$ & $ 5.48 \pm 1.18$ & $0.78\pm0.04$& $ 43.41 \pm 4.93$ & $ 3.85 \pm 1.03$ & $0.92\pm0.02$\\
        $7$ & $|\eta|<2.0$ & $0.94$ & $ 0.37 \pm 0.02$ & $ 10.63 \pm 0.43$ & $ 2.91 \pm 0.22 $ & $0.79\pm0.01$& $36.84 \pm 1.41 $ & $2.51 \pm 0.20$ & $0.94\pm0.01$\\
        &  $|\eta|<2.4 $ & $0.95$ & $ 0.35 \pm 0.03 $ & $12.73 \pm 0.68$ & $ 3.17 \pm 0.40 $ & $0.80\pm0.02$& $43.05 \pm 2.21$ & $ 2.47 \pm 0.27$ & $0.95\pm0.01$\\
        &  $|\eta|< 3.0 $ & $ 0.94$ & $0.32 \pm 0.03 $ & $ 15.55 \pm 0.94 $ & $ 3.52 \pm 0.58 $ & $0.82\pm0.03$& $ 52.08 \pm 2.93 $ & $ 2.45 \pm 0.29 $ & $0.96\pm0.01$\\
          &  $|\eta|< 3.4 $ & $0.94$ & $ 0.31 \pm 0.03$ & $ 17.43 \pm 1.05$ & $ 3.68 \pm 0.65$ & $0.83\pm0.03$ & $ 57.38 \pm 3.33$ & $ 2.43 \pm 0.30$ & $0.96\pm0.01$\\
&  $-3.4<\eta< +5.0 $ & $0.94$ & $0.30 \pm 0.03$ & $ 20.74 \pm 1.28$ & $ 4.18 \pm 0.77$& $0.83\pm0.03$ & $ 66.40 \pm 4.08$ & $ 2.54 \pm 0.34$ & $0.96\pm0.01$\\
$8$ & $|\eta|<2.0$ & $0.94$ & $ 0.45 \pm 0.03$ & $ 12.37 \pm 0.79$ & $ 2.38 \pm 0.15$ & $0.84\pm0.01$& $ 41.16 \pm 2.01$ & $ 2.93 \pm 0.29$ & $0.93\pm0.01$\\
&  $|\eta|<2.4 $ & $0.94$ & $ 0.37 \pm 0.04$ & $ 13.71 \pm 1.10$ & $ 2.75 \pm 0.34$ & $0.83\pm0.02$ & $ 45.73 \pm 3.15$ & $ 2.56 \pm 0.37$ & $0.95\pm0.01$\\
&$|\eta|< 3.0$ & $0.94$ & $ 0.26 \pm 0.03$ & $ 15.50 \pm 0.99$ & $ 3.63 \pm 0.78$ & $0.81\pm0.03$& $ 51.58 \pm 3.25$ & $ 2.19 \pm 0.29$ & $0.96\pm0.01$\\
&  $|\eta|< 3.4 $ & $0.93$ & $ 0.26 \pm 0.03$ & $ 17.57 \pm 0.91$ & $ 3.84 \pm 0.75$ & $0.82\pm0.03$& $ 57.55 \pm 3.17$ & $ 2.22 \pm 0.27$ & $0.96\pm0.01$\\
&  $-3.4<\eta< +5.0 $ & $0.94$ & $ 0.17 \pm 0.03$ & $ 20.11 \pm 0.56$ & $ 6.65 \pm 2.45$ & $0.75\pm0.07$& $ 62.60 \pm 3.03$ & $ 2.00 \pm 0.21$ & $0.97\pm0.01$
\end{tabularx}
\end{subtable}
\begin{subtable}{\textwidth}
\caption*{INEL$>0$}\label{table: INEL>0}
\renewcommand{\arraystretch}{1.4}%
\begin{tabularx}{\textwidth}{l c c*{8}{Y}}
$\sqrt{s}$  & $\eta$ range & $\lambda \ (\pm 0.01)$ & $\alpha$ & $\langle n \rangle_1$ & $k_1$ & $\langle p \rangle_1$ & $\langle n \rangle_2$ & $k_2$ & $\langle p \rangle_2$\\ \hline \\
 $0.9$ & $|\eta|<2.0$& $1.00$ & $0.56 \pm 0.04$ & $ 10.54 \pm 0.69$ & $ 3.31 \pm 0.21$ & $0.76\pm0.02$& $ 25.78 \pm 2.04$ & $ 4.08 \pm 0.65$ & $	
0.86\pm 0.02$\\
 &$|\eta|<2.4$ & $1.00$ & $ 0.45 \pm 0.06$ & $ 11.84 \pm 1.11$ & $ 3.37 \pm 0.37$ & $0.77\pm 0.02$& $ 28.64 \pm 2.93$ & $ 3.49 \pm 0.73$ & $0.89\pm0.02$\\
 &$|\eta|< 3.0$ & $0.97$ &  $ 0.23 \pm 0.07$ & $ 13.42 \pm 0.56 $ & $6.30 \pm 3.33$ & $0.68 \pm 0.12$& $ 30.77 \pm 3.00$ & $ 2.71 \pm 0.55$ & $0.92\pm0.02$\\
 &  $|\eta|< 3.4 $ & $0.95$ & $ 0.33 \pm 0.05$ & $ 15.76 \pm 0.90$ & $ 4.80 \pm 0.98$ & $0.77 \pm 0.04$ & $ 36.74 \pm 2.96$ & $ 3.22 \pm 0.55$ & $0.92\pm 0.01$\\
 &  $-3.4<\eta< +5.0 $ & $0.92$ & $0.41 \pm 0.05 $ & $19.37 \pm 1.10 $ & $4.77 \pm 0.54$ & $0.80\pm 0.02$& $ 44.72 \pm 3.20$ & $ 4.07 \pm 0.68$ & $0.92\pm 0.01$\\
  $7$ & $|\eta|<2.0$& $1.00$& $ 0.37 \pm 0.02$& $ 10.88 \pm 0.42$& $ 3.03 \pm 0.22$& $0.78 \pm 0.01$& $ 36.77 \pm 1.41$& $ 2.53 \pm 0.21$& $0.94\pm0.01$\\
   &$|\eta|< 2.4$ & $1.00$ & $ 0.35 \pm 0.03$ & $ 12.87 \pm 0.69$ & $ 3.21 \pm 0.41$ & $0.80\pm0.02$ & $ 42.83 \pm 2.27$ & $ 2.48 \pm 0.27$ & $0.95\pm 0.01$\\
    &$|\eta|< 3.0$ & $0.98$ & $ 0.32 \pm 0.03$ & $ 15.58 \pm 1.00$ & $ 3.56 \pm 0.67$ & $0.81\pm0.03$& $ 51.50 \pm 3.09 $ & $2.44 \pm 0.30$ & $0.95\pm 0.01$\\
    &  $|\eta|< 3.4 $ & $0.96$ & $0.31 \pm 0.03$ & $ 17.53 \pm 1.15$ & $ 3.63 \pm 0.69$ & $0.83 \pm 0.03$& $ 57.02 \pm 3.52$ & $ 2.45 \pm 0.32$ & $0.96\pm 0.01$\\
     &  $-3.4<\eta< +5.0 $ & $0.94$ & $ 0.31 \pm 0.03$ & $ 20.70 \pm 1.37$ & $ 3.94 \pm 0.72$ & $0.84 \pm 0.03$& $ 66.27 \pm 4.07$ & $ 2.60 \pm 0.34$ & $0.96 \pm 0.01$\\
      $8$ & $|\eta|<2.0$& $1.00$& $0.43 \pm 0.02 $& $10.83 \pm 0.52$& $ 2.54 \pm 0.18 $& $0.81\pm 0.01$& $38.14 \pm 1.48 $& $2.73 \pm 0.21$& $0.93\pm 0.01$\\
      &$|\eta|< 2.4$ & $0.99$ & $0.37 \pm 0.03$ & $ 12.45 \pm 0.81$ & $ 2.93 \pm 0.39$ & $0.81\pm 0.02$& $ 43.33 \pm 2.62$ & $ 2.47 \pm 0.31$ & $0.95\pm0.01$\\
      &$|\eta|< 3.0$ & $0.97$ & $0.28 \pm 0.03$ & $ 14.61 \pm 0.77$ & $ 3.73 \pm 0.80$ & $0.80\pm 0.04$& $ 50.01 \pm 2.90$ & $ 2.19 \pm 0.27$ & $0.96\pm 0.01$\\
      &  $|\eta|< 3.4 $ & $0.95$ & $ 0.30 \pm 0.03$ & $ 16.79 \pm 0.80$ & $ 3.61 \pm 0.60$ & $0.82\pm 0.03$& $ 56.72 \pm 2.97$ & $ 2.28 \pm 0.26$ & $0.96\pm 0.01$\\
      &  $-3.4<\eta< +5.0 $ & $0.93$ & $0.28 \pm 0.03$ & $ 20.00 \pm 0.92$ & $ 4.08 \pm 0.77$ & $0.83\pm 0.03$& $ 65.51 \pm 3.77$ & $ 2.32 \pm 0.30$ & $0.97\pm 0.01$
\end{tabularx}
\end{subtable}
\end{table}
\begin{table}
\ContinuedFloat
\centering
\begin{subtable}{\textwidth}
\caption*{INEL}\label{table: INEL}
\renewcommand{\arraystretch}{1.4}%
\begin{tabularx}{\textwidth}{l c c*{8}{Y}}
$\sqrt{s}$  & $\eta$ range & $\lambda \ (\pm 0.01)$ & $\alpha$ & $\langle n \rangle_1$ & $k_1$ & $\langle p \rangle_1$ & $\langle n \rangle_2$ & $k_2$ & $\langle p \rangle_2$\\ \hline \\
 $0.9$ & $|\eta|<2.0$& $0.81$&$ 0.64 \pm 0.05$ & $10.44 \pm 0.97$ &  $ 2.69 \pm 0.20$ & $0.80\pm0.02$ & $ 27.03 \pm 2.71$ & $ 4.45 \pm 0.91$ & $0.86\pm0.03$ \\
 &$|\eta|<2.4$ & $0.81$& $0.33 \pm 0.05$& $ 10.12 \pm 0.61$& $ 4.00 \pm 0.93 $& $0.72\pm0.05$& $25.48 \pm 2.11$& $ 2.80 \pm 0.47$ & $0.90\pm0.02$ \\
 &$|\eta|< 3.0$ &$0.80$& $ 0.31 \pm 0.05$& $ 12.91 \pm 0.66$& $ 4.83 \pm 1.39$& $0.73\pm0.06$ & $ 31.45 \pm 2.67$& $ 2.83 \pm 0.50$& $0.92\pm0.01$ \\
 &  $|\eta|< 3.4 $ &$ 0.80$ &$0.40 \pm 0.05$ &$ 15.30 \pm 0.95$ &$ 4.10 \pm 0.63$ &$0.79\pm0.03$&$ 37.39 \pm 2.97 $ &$3.33 \pm 0.56$ &$0.92\pm0.01$ \\
 &  $-3.4<\eta< +5.0 $ &$0.80$ &$ 0.46 \pm 0.04$ &$ 18.83 \pm 1.08$ &$ 4.32 \pm 0.45$ &$0.81\pm0.02$&$ 45.16 \pm 3.18$ &$ 4.16 \pm 0.69$ &$0.92\pm0.01$ \\
  $7$ & $|\eta|<2.0$ &$0.79$ &$ 0.43 \pm 0.02 $ &$10.42 \pm 0.49$ &$ 2.43 \pm 0.17$ &$0.81\pm0.01$&$ 37.54 \pm 1.56$ &$ 2.63 \pm 0.23$ &$0.93\pm0.01$ \\
   &$|\eta|< 2.4$ &$0.79$ &$ 0.38 \pm 0.03$ &$ 12.12 \pm 0.71 $ &$2.79 \pm 0.38 $ &$0.81\pm0.02$&$42.87 \pm 2.36 $ &$2.48 \pm 0.28$ &$0.95\pm0.01$ \\
    &$|\eta|< 3.0$ &$0.78$ &$ 0.35 \pm 0.03$ &$ 14.75 \pm 1.01 $ &$3.13 \pm 0.63$ &$0.82\pm0.03$ &$ 51.55 \pm 3.28$ &$ 2.45 \pm 0.32$ &$0.95\pm0.01$ \\
    &  $|\eta|< 3.4 $ &$0.78$ &$ 0.35 \pm 0.03 $ &$16.69 \pm 1.14$ &$ 3.15 \pm 0.57$ &$0.84\pm0.03$ &$ 57.31 \pm 3.63$ &$ 2.48 \pm 0.33$ &$0.96\pm0.01$ \\
     &  $-3.4<\eta< +5.0 $ &$0.78$ &$ 0.33 \pm 0.03 $ &$19.54 \pm 1.21$ &$ 3.75 \pm 0.77$ &$0.84\pm0.03$&$ 65.47 \pm 4.01$ &$ 2.53 \pm 0.33$ &$0.96\pm0.01$ \\
      $8$ & $|\eta|<2.0$ &$ 0.80$ &$ 0.52 \pm 0.03 $ &$10.87 \pm 0.84 $ &$1.93 \pm 0.25$ &$0.85\pm0.02$&$ 40.01 \pm 2.39$ &$ 2.98 \pm 0.35$ &$0.93\pm0.01$ \\
      &$|\eta|< 2.4$ &$0.79$ &$ 0.42 \pm 0.04 $ &$12.08 \pm 0.98$ &$ 2.37 \pm 0.38$ &$0.84\pm0.02$ &$ 44.39 \pm 3.12$ &$ 2.58 \pm 0.37$ &$0.95\pm0.01$ \\
      &$|\eta|< 3.0$ &$0.78$ &$0.32 \pm 0.03$ &$ 13.94 \pm 0.85$ &$ 3.10 \pm 0.64 $ &$0.82\pm0.03$&$50.50 \pm 3.16$ &$ 2.23 \pm 0.29$ &$0.96\pm0.01$ \\
      &  $|\eta|< 3.4 $ &$0.78$ &$ 0.34 \pm 0.03$ &$ 16.03 \pm 0.86$ &$ 3.08 \pm 0.49$ &$0.84\pm0.02$&$ 57.14 \pm 3.15$ &$ 2.32 \pm 0.28$ &$0.96\pm0.01$ \\
      &  $-3.4<\eta< +5.0 $ &$0.78$ &$ 0.30 \pm 0.03$ &$ 19.03 \pm 0.86 $ &$3.78 \pm 0.74$ &$0.83\pm0.02$&$ 65.08 \pm 3.74$ &$ 2.29 \pm 0.29$ &$0.97\pm0.01$
\end{tabularx}
\end{subtable}
\end{table}
\newpage
\appendix
\setcounter{equation}{0}
\renewcommand{\theequation}{A\arabic{equation}}

\section{Classical Mean Square Dynamics in a Canonical Ensemble}\label{appendix A}
The \textit{heat kernel} is the fundamental solution to the heat equation $\text{d} K/\text{d} t = D\nabla^2 K$ on a particular domain with a given set of boundary conditions, and in the case of a single spatial dimension for $t>0$, it gives the probability density function of a particle being displaced from a position $x_0$ to $x$ in a time interval $\Delta t=t-t_0$. It can be mathematically shown that the heat kernel in one dimension has the following form:
\begin{align}
K(x,t|x_0,t_0)\ \text{d} x=\sqrt{\frac{1}{4\pi D  (t-t_0)}}\exp\left[-\frac{(x-x_0)^2}{4D (t-t_0)}\right] \text{d} x,
\end{align}
where $D$ is the diffusion coefficient \cite{einstein1905motion} \cite{grigoryan2009heat}. By its Markov property, an arbitrary heat kernel can be defined as the product of kernels, defined over successive intervals of displacement and time, such that
\begin{align}
K(x_c,t_c|x_0,t_0)= K(x_c,t_c|x_b,t_b)  K(x_b,t_b|x_a,t_a),\label{eq}
\end{align}
where $x_c>x_b>x_a$ and $t_c>t_b>t_a$. Therefore, without loss of generality, one can define a kernel for a displacement in an infinitesimal time interval, from which arbitrary kernels can be defined. As such, we consider the limiting case as $\Delta t \to 0$:
\begin{align}
\lim_{\Delta t \to 0} K(x,t | x_0, t_0)\ \text{d} x &=\lim_{\Delta t \to 0} \sqrt{\frac{1}{4\pi D  (t-t_0)}}\exp\left[-\frac{(x-x_0)^2}{4D (t-t_0)}\right] \text{d} x.
\end{align}
Letting $x_0=t_0=0$ and relabeling $x=\Delta x$ and $t=\Delta t$, we have
\begin{align}
\lim_{\Delta t \to 0} K(\Delta x,\Delta t|0,0)\ \text{d}(\Delta  x) &=\lim_{\Delta t \to 0} \sqrt{\frac{1}{4\pi D \Delta t}}\exp\left[-\frac{1}{4D}\frac{(\Delta x)^2}{\Delta t}\right] \text{d} (\Delta  x).
\end{align}
In order to make the connection with the Maxwell-Boltzmann (MB) distribution, we introduce a measure on the infinitesimal space of position and time, and we interpret $K(x,t|x_0,t_0)$ as a probability density function in this space. By replacing the limit with the formal substitution $\Delta t \to \text{d}t$ and $\Delta x \to \text{d}x$, we can write
\begin{align}
K(\text{d}x,\text{d}t|0,0) \ \text{d}(\text{d} x) &=\sqrt{\frac{1}{4\pi D \text{d}t}}\exp\left[-\frac{1}{4D}\frac{\text{d} x^2}{\text{d} t}\right] \text{d} (\text{d} x). \label{eq: hq}
\end{align}
We now turn our attention to the MB distribution. For a canonical ensemble of particles of mass $m$ and temperature $T$, the MB distribution gives the probability of finding a particle with an instantaneous velocity $v$. In terms of a single degree of freedom in an element of the velocity phase space, the distribution takes the form
\begin{align}
f(v) \ \text{d} v &=\sqrt{\frac{m}{2\pi k_BT}}\exp\left[-\frac{mv^2}{2k_BT}\right] \ \text{d} v,
\end{align}
where the mean square speed is $\langle v^2 \rangle= k_BT/m$ and $k_B$ is the Boltzmann constant. Making the substitution $v= \text{d} x /\text{d} t$ and introducing factors of $\text{d} t$, we have
\begin{align}
f(v) \ \text{d} v &=\sqrt{\frac{m}{2\pi k_BT\text{d} t^2}}\exp\left[-\frac{m}{2k_BT}\left(\frac{\text{d} x}{\text{d} t}\right)^2\right] \ \text{d} t \ \text{d} v.
\end{align}
By virtue of the fact that $\text{d} v / \text{d} t = \text{d}^2 x  / \text{d} t^2$, such that $\text{d} t \ \text{d} v = \text{d}^2 x = \text{d}(\text{d} x)$, we find
\begin{align}
f(v) \ \text{d} v &=\sqrt{\frac{m}{2\pi k_BT\text{d} t^2}}\exp\left[-\frac{m}{2k_BT\text{d} t}\frac{\text{d} x^2}{\text{d} t}\right] \ \text{d}(\text{d} x), \label{eq: mb}
\end{align}
such that we can equate Eq. (\ref{eq: hq}) with (\ref{eq: mb}). Given this equality, the diffusion coefficient can be expressed as
\begin{align}
D&= \lim_{\Delta t \to 0} \ \frac{k_BT \Delta t} {2m}.
\end{align}
This resembles the Einstein relation \cite{einstein1905motion}, $D=\mu k_B T$, for an ensemble of density $\rho$ in the presence of an external potential $U(x)$, whereby the mobility, $\mu : [s \cdot kg^{-1}]$, is related to the particle drift current $j=-\mu \rho (\text{d} U / \text{d} x)$. However, in the absence of an external potential or any particle interactions---as is considered in the present context---there is no drift current, and the mobility is shown to be replaced by the ratio $\Delta t /(2m)$. It follows that the diffusion coefficient can be related to the mean square speed as
\begin{align}
2D&= \lim_{\Delta t \to 0} \ \langle v^2 \rangle \Delta t.
\end{align}
Moreover, as the mean square displacement in one dimension satisfies the identity \cite{einstein1905motion}
\begin{align}
\langle d^2 \rangle = \langle |x(t) - x_0|^2\rangle = 2D \Delta t,
\end{align}
we find, in the limit under consideration, that
\begin{align}
 \langle v^2 \rangle = \lim_{\Delta t \to 0} \frac{\langle d^2 \rangle}{(\Delta t)^2}. 
\end{align}
By applying L'Hospital's rule, we can confirm the limiting expressions of $2D=\textstyle{\lim_{\Delta t \to 0}} \ \langle d^2 \rangle / \Delta t$ and $2D=\textstyle{\lim_{\Delta t \to 0}} \ \langle v^2 \rangle \Delta t$ are in agreement. As these relations hold for an infinitesimal time-step, by the Markov property of kernels, they can be shown to hold for arbitrarily large time intervals, as well. Integrating over all possible displacements for each kernel factor, one can write the resulting kernel as 
\begin{align}
K(x,t|x_0,t_0) &= \lim_{\Delta t \to 0} \int_{-\infty}^{\infty} \dots \int_{-\infty}^{\infty} \left(\frac{1}{4\pi D \Delta t}\right)^{n/2}\prod_{j=1}^n \exp\left[ -\frac{(x_j-x_{j-1})^2}{4D \Delta t} \right] \ \text{d} x_{n-1} \dots \text{d} x,
\end{align}
which is precisely the Euclidean path integral definition of the kernel \cite{einstein1905motion} \cite{feynman2010quantum}. In each of the $n$ infinitesimal intervals, the identity  $\langle v^2 \rangle= \lim_{\Delta t \to 0} \langle d^2 \rangle/(\Delta t)^2$ holds. Hence, the identity holds for a finite time interval in general:
\begin{align}
\langle v^2 \rangle = \frac{\langle d^2 \rangle}{(\Delta t)^2}.
\end{align}
The generalization to higher dimensions follows naturally from these definitions.

\setcounter{equation}{0}
\renewcommand{\theequation}{B\arabic{equation}}
\section{Relativistic Mean Square Dynamics in a Canonical Ensemble}\label{appendix B}

The following is a relativistic generalization of the results in Appendix \ref{appendix A}. In a given inertial reference frame, let each measurement of the particles in a canonical ensemble be evaluated over a common interval of proper time $\Delta \tau = \tau - \tau_0$. The mean square displacement of the ensemble is parametrized with respect to $\tau$, such that in the case of a single spatial dimension,
\begin{align}
\langle d^2 \rangle \equiv \langle |x(\tau)-x(\tau_0)|^2 \rangle.
\end{align}
Furthermore, we me must introduce a mean square temporal displacement, as well, which is parametrized analogously to $\langle d^2 \rangle$, such that
\begin{align}
\langle t^2 \rangle \equiv \langle |t(\tau)-t(\tau_0)|^2 \rangle.
\end{align}
Each $i$th particle has a particular mean velocity $\langle v_i \rangle$---defined as the ratio of its displacement $\Delta x_i = x_i(\tau)-x_i(\tau_0)$ over a particular interval of absolute time $\Delta t_i = t_i(\tau)-t_i(\tau_0)$, such that $\langle v_i \rangle= \Delta x_i / \Delta t_i$---and must satisify the additional constraint that its displacement vector $\vec X_i = (\Delta t_i, \Delta x_i)$ lies on the hypersurface
\begin{align}
\mathcal{S} \quad : \quad c^2(\Delta t)^2 - (\Delta x)^2 = c^2(\Delta \tau)^2. \label{eq: hypersurface cond}
\end{align}
The mean relativistic velocity of each $i$th particle is thus $\langle u_i \rangle = \Delta x_i /\Delta  \tau$. As each particle's velocity is a ratio of the components of its spacetime displacement vector, and vector addition is a component-wise operation, it follows that the mean velocity corresponding to an average spacetime displacement of several particles is defined by the mediant of the mean particle velocities. That is, for $N$ particles, the displacement vectors $\vec X_1, \vec X_2, \dots, \vec X_N$ have an average displacement vector $\langle\vec X \rangle\equiv (\vec X_1 + \vec X_2 + \dots + \vec X_N)/N = ((\Delta t_1 + \Delta t_2+ \dots +\Delta t_N)/N , (\Delta x_1 + \Delta x_2 + \dots + \Delta x_N)/N)$, such that
\begin{align}
\langle v \rangle \equiv \frac{\frac{1}{N}(\Delta x_1 + \Delta x_2 + \dots + \Delta x_N) } {\frac{1}{N}(\Delta t_1 + \Delta t_2 + \dots + \Delta t_N)} = \frac{\Delta x_1 + \Delta x_2 + \dots + \Delta x_N} {\Delta t_1 + \Delta t_2 + \dots + \Delta t_N}.
\end{align}
Likewise, the mean square speed $\langle v^2 \rangle$  is defined as the mediant of the squared mean velocities, $\langle v_1 \rangle^2, \langle v_2 \rangle^2, \dots, \langle v_N \rangle^2$, such that
\begin{align}
\langle v^2 \rangle  = \frac{\frac{1}{N}\left[(\Delta x_1)^2 + (\Delta x_2)^2 + \dots +(\Delta x_N)^2\right]}{\frac{1}{N}\left[(\Delta t_1)^2 + (\Delta t_2)^2 + \dots + (\Delta t_N)^2\right]} =  \frac{(\Delta x_1)^2 + (\Delta x_2)^2 + \dots +(\Delta x_N)^2}{(\Delta t_1)^2 + (\Delta t_2)^2 + \dots + (\Delta t_N)^2}.
\end{align}
Notice, however, that $\langle v \rangle$ does \textit{not} satisfy the hypersurface condition of Eq. (\ref{eq: hypersurface cond}), but the root mean square speed $\tilde v = \sqrt{\langle v^2 \rangle}$ does, as
\begin{align}
c^2 \langle t^2 \rangle - \langle d^2 \rangle &=\frac{c^2}{N}\left[(\Delta t_1)^2 + (\Delta t_2)^2 + \dots + (\Delta t_N)^2\right] - \frac{1}{N}\left[(\Delta x_1)^2 + (\Delta x_2)^2 + \dots +(\Delta x_N)^2\right]\\
&= \frac{1}{N}\left[c^2(\Delta t_1)^2 - (\Delta x_1)^2\right] + \frac{1}{N}\left[c^2(\Delta t_2)^2 - (\Delta x_2)^2\right] + \dots + \frac{1}{N}\left[c^2(\Delta t_N)^2 - (\Delta x_N)^2\right]\\
&= c^2 (\Delta \tau)^2,
\end{align}
and therefore, the inertial observer identifies the absolute mean square speed as
\begin{align}
\langle v^2 \rangle  = \frac{\langle d^2 \rangle}{\langle t^2 \rangle},
\end{align}
and the relativistic mean square speed as 
\begin{align}
\langle u^2 \rangle = \frac{\langle d^2 \rangle}{(\Delta \tau)^2}
\end{align}
with a mean square Lorentz factor 
\begin{align}
\langle \gamma^2 \rangle = \frac{\langle t^2 \rangle}{(\Delta \tau)^2}.
\end{align}
The generalization to higher dimensions follows naturally.

\setcounter{equation}{0}
\renewcommand{\theequation}{C\arabic{equation}}

\section{Negative Binomial Distribution}\label{sec: NBD}
The negative binomial distribution $\text{NB}(k, \langle p \rangle)$ is defined by a probability mass function $f_{\text{\tiny{NB}}}(n; k, \langle p \rangle)$ that models the number of $n$ successes in a sequence of independent and identically-distributed Bernoulli trials of average probability $\langle p \rangle$, given that a specific (non-random) number of failures $k$ occurs \cite{klugman2012loss}: \begin{align}\label{eq: NBD Appen}
X \sim \text{NB}(k, \langle p \rangle) \quad : \quad f_{\small{NB}}(n; k, \langle p \rangle) \equiv  \Pr(n) = \binom{n+k-1}{n}(1-\langle p \rangle)^{k} {\langle p \rangle}^n.
\end{align}
To eliminate the ambiguity in terminology between the probability $\Pr(X=n)$ and probability $\langle p \rangle$, we will refer to the latter as the mean \textit{chance} of an observation. Averaged over many statistical experiments, the mean number of trials $\langle \mathcal{N} \rangle$ is defined by the NBD in terms of the average count $\langle n \rangle$, as $\langle \mathcal{N} \rangle = \langle n \rangle / \langle p \rangle$; or in terms of $k$, as $\langle \mathcal{N} \rangle = \langle n \rangle + k$. We will continue to denote the chance $\langle p \rangle$ with angle brackets as a reminder that it is also an average quantity. As the value of $k$ is constant in all experiments, it is not interpreted as an average.\\

To briefly review the relationship of the binomial distribution to the NBD, suppose $n$ is a random, binomially-distributed variable with parameters ${\mathcal{N}}$ and $\langle p \rangle$. If $\langle p \rangle+\langle q \rangle=1$, for $\langle p \rangle, \langle q \rangle \geq 0$, then
\begin{align}
	1=1^{\mathcal{N}}=(\langle p \rangle+\langle q \rangle)^{\mathcal{N}}= \sum_{k=0}^{\infty}\binom{{\mathcal{N}}}{n}  \langle q \rangle^{{\mathcal{N}}-n}\langle p \rangle^{n},
\end{align}
and when generalized for a real-valued ${\mathcal{N}}$, Newton's binomial theorem gives
\begin{align}
	\binom{{\mathcal{N}}}{n} = \frac{{\mathcal{N}}({\mathcal{N}}-1)({\mathcal{N}}-2)\dots({\mathcal{N}}-n+1)}{n!}.
\end{align}
Making the substitution ${\mathcal{N}}_\ast={\mathcal{N}}-1$, the above becomes
\begin{align}
	\binom{{\mathcal{N}}_\ast}{n} = \frac{(n+k-1)(n+k-2)(n+k-3)\dots k}{n!} = (-1)^n\frac{-k(-k-1)(-k-2)\dots (-k-n+1)}{n!}= (-1)^n\binom{-k}{n}.
\end{align}
Therefore, one can write
\begin{align}
1=\langle q \rangle^{k} \langle q \rangle^{-k}  &= \langle q \rangle^{k} (1-\langle p \rangle)^{-k}  = \sum_{n=0}^{\infty}\binom{-k}{n}\langle q \rangle^{k} (-\langle p \rangle)^{n} = \sum_{n=0}^{\infty}\binom{n+k-1}{n} (1-\langle p \rangle)^{k} \langle p \rangle^{n},
\end{align}
which is the negative binomial distribution defined in Eq. (\ref {eq: NBD Appen}) \cite{feller2008introduction}. Therefore, the NBD can also be expressed as
\begin{align} \label{eq: negative NBD coefficient form}
 f_{\small{NB}}(n; k, \langle p \rangle)=	\binom{-k}{n}(1-\langle p \rangle)^{k} (-\langle p \rangle)^{n}. 
\end{align}

The NBD is a robust alternative to the Poisson distribution as it generalizes the latter in the limit as $k \to \infty$ and $\langle p \rangle \to 0$. In contrast to the Poisson distribution, the variance of the NBD can differ from its mean and is therefore useful in studying overdispersed count data \cite{feller2008introduction}.

As its name suggests, the NBD permits negative values of $k$ and hence $\langle \mathcal{N} \rangle$, as well. In fact, it can be easily shown that $k, \langle \mathcal{N} \rangle \in \mathbbm{R}$, therefore permitting the use of the gamma function in the binomial coefficient, such that 
\begin{align}
f_{\small{NB}}(n; k, \langle p \rangle) = \frac{\Gamma(n+k)}{n!\Gamma(k)} (1-\langle p \rangle)^{k} {\langle p \rangle}^n.\label{eq: NBD expanded}
\end{align}

\end{document}